\documentclass[12pt,a4paper]{article}
\usepackage{latexsym}
\usepackage{amsfonts}
\usepackage{multicol}
\usepackage[dvips]{graphicx}
\usepackage{fixltx2e}

\newcommand{\danger}[1]{\textbf{#1}}

\input{epsf}             % encapsulated Postscript figures
\begin{document}

\title{\danger{Towards a spin foam model description of black hole entropy}}
\author{\centerline{\danger{J. Manuel Garc\'\i a-Islas \footnote{
e-mail: jmgislas@leibniz.iimas.unam.mx}}}  \\
Instituto de Investigaciones en Matem\'aticas Aplicadas y en Sistemas \\ 
Universidad Nacional Aut\'onoma de M\'exico, UNAM \\
A. Postal 20-726, 01000, M\'exico DF, M\'exico\\}

\maketitle

\begin{abstract}
We propose a way to describe the origin of black hole entropy
in the spin foam models of quantum gravity. This stimulates a new way to study the relation
of spin foam models and loop quantum gravity.
\end{abstract}

\begin{multicols}{2}

\section{Introduction}

Derivation of the Bekenstein-Hawking black hole entropy \cite{b}, \cite{h}
from a quantum theory of gravity is one of the most important problems of physics. 
It is a must for any theory of quantum gravity to give a satisfactory statistical description.
Here we propose a way to describe black hole entropy in a theory of quantum gravity known as
spin foam models. Spin foam models are the covariant version of loop quantum gravity. 

Loop quantum gravity is a very promising theory  which has studied the statistical  
description of black hole entropy where seminal ideas come from \cite{kk}, \cite{cr}.
Later, the problem has been developed deeper following different approaches for the counting of the degrees of freedom \cite{abck}, \cite{dl}, \cite{cpb}, \cite{t}, \cite{abpbv}, \cite{tt}. The problem has even
been extended to horizon surface of higher genus \cite{kbd}.

The problem of giving a statistical description of the entropy in terms of
spin foam models has been considered for the 
three dimensional BTZ black hole only \cite{gi}. Here we present a proposal to
study the entropy 
of a four dimensional spherically symmetric black hole in terms of spin foam models. The study of
back hole entropy could be a
way to relate loop quantum gravity to spin foam models or at least it can
give more evidence of the relation.

We divide this letter as follows: In section 2 we describe very briefly the idea behind the statistical description
of black hole entropy in loop quantum gravity. In section 3 we propose a way to study the description
of black hole entropy
in spin foam models.  

\section{Black hole entropy in loop quantum gravity}

Let us consider the basic ideas behind black hole entropy in loop quantum gravity. 
Consider the surface of a Schwarzschild black hole. In loop quantum gravity its geometry is determined by the punctures of the edges of a spin network embedded in three dimensional space with the surface. Label the edges of
the spin network by $j_{i} $ which are half integers, that is, irreducible representations
of the group $SU(2)$.
Suppose the spin network punctures the surface in $n$ isolated points, and in a non-degenerate way.
The total area of the surface is given by the eigenvalues of the area operator $A$

\begin{equation}
A= 8 \pi \gamma {\l_{p}}^2 \sum_{i=1}^{n} \sqrt{j_{i}(j_{i}+1)}
\end{equation}
where $\l_{p}$ is the Planck length and $\gamma$ is a parameter known by the name of Immirzi. 

The entropy of the black hole is given by the logarithm of the number $N$ of states which account for a fixed area of the surface 

\begin{equation}
S= \log N
\end{equation}
The number  $N$ of these states
has being considered in 
\cite{abck}, \cite{dl}, \cite{cpb}, \cite{t}, \cite{abpbv}, \cite{tt}; however the states which account for
the entropy vary in opinions and we have various possibilities. 

The question is whether we can get a counting related to the one in loop quantum gravity 
by using the spin foam models of quantum gravity. 

\section{Black hole entropy from spin foam models}

In this section we propose a way to compute the entropy of a spherically symmetric black hole
in spin foam models of quantum gravity.
In this letter we will just give a description of our idea and the important calculations 
will appear in a future paper.

\begin{figure*}[!]
\centering
\includegraphics[width=0.7\textwidth]{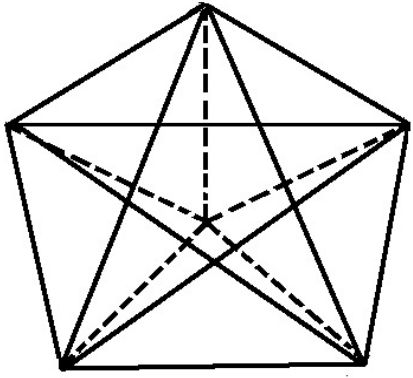}
\caption{4-simplex subdivision}
\end{figure*}

In spin foam models we think of space-time as discrete. It is thought for instance as a  2-complex 
dual to a triangulation of the space-time manifold, or as the triangulation itself. In \cite{gi} we considered the entropy of
BTZ black hole from a spin foam model viewpoint. Here we do something similar. 
Consider a spherical symmetric black hole such as Schwarzschild. Take a three dimensional
subspace such as a Cauchy surface in which the spherical surface of the horizon is embedded. 
This is, we take $S^2$ embedded in $R^3$. 
Decompose $S^2$ as a 2-complex.
Such decomposition will be of course part of the 2-complex which represents the whole space-time.

Consider a 2-complex decomposition of the horizon surface $S^2$, one which is dual to a non-degenerate triangulation.
We can think that the non-degenerate triangulation of the sphere $S^2$ was obtained 
by subdivisions of 4-simpleces  as figure $[1]$. 
The edges of the graph of figure $[1]$ correspond to faces of the dual 2-complex.
We can also think that three of the dotted lines of the figure correspond to faces of the horizon surface. 

All the remaining faces of the horizon surface will be part of other
subdivided 4-simpleces.    

Consider a spin foam state sum where faces of the 2-complex are labelled by spins
which are irreducible representations of a group or a quantum group. 
The sum is over admissible irreducible
representations of the product of amplitudes given to faces and vertices. 

We will take the spins labelling the faces of the horizon surface of the black hole as fixed.
Denote the horizon surface as $\mathcal{O}$.
Consider the state sum

\begin{equation}
Z(M,\mathcal{O}) = \sum_{S \mid
\mathcal{O}} \prod_{faces}dim_{q}(j)  
\prod_{vertices} A_v
\end{equation}
where $M$ is the 2-complex which represents the space-time, $A_v$ is the amplitude of vertices, 
$S\mid\mathcal{O}$ means that the irreducible representations which label the horizon faces
are not summed over. Clearly the state sum is a function of the labels of the horizon.

In the dictionary of loop quantum gravity versus spin foam models we propose that the punctures
of the spin network with the horizon surface correspond in spin foam models to this description of having fixed spins at the horizon in the state sum.

We define the entropy of the black hole as

\begin{equation}
\log \bigg(\frac{Z(M,\mathcal{O})}{Z(M)} \bigg)
\end{equation}
where $Z(M)$ is the state sum of the space-time manifold without fixed spins at the horizon.
The idea now is to prove that our definition of the entropy is related to the 
entropy computed in loop quantum gravity. The important question now is
which state sum will get the entropy of loop quantum gravity. Alternatively we ask, which spin foam
model is the one which will lead us to an entropy related to the one of loop quantum gravity.
We have the BC model \cite{bc}, and two new models which are more promising
\cite{fk},\cite{epc}.

It could be very interesting to study formula $(4)$ for these models to see whether it is related 
to the calculation of black hole entropy in loop quantum gravity. In a future paper we will compute
formula $(4)$ for the case of Riemannian BC model with cosmological constant,
which means that we will use the quantum group $SO_{q}(4)$.

\end{multicols}

\end{document}